\documentclass[proceedings, preprint]{rmaa}

\usepackage{paralist}

\usepackage{psfrag,color}

\SetYear{2009}
\SetConfTitle{A Long Walk Through Astronomy}

\title{AGN and starburst in bright seyfert galaxies: from IR photometry to IR spectroscopy} 

\author{
  Luigi Spinoglio,\altaffilmark{1} 
  Silvia Tommasin,\altaffilmark{1}
  and Matthew A. Malkan \altaffilmark{2}}

\altaffiltext{1}{Istituto di Fisica dello Spazio Interplanetario,
  INAF, Via Fosso del Cavaliere 100, I-00133, Roma,
  Italy (luigi.spinoglio, silvia.tommasin@ifsi-roma.inaf.it).}

\altaffiltext{2}{Astronomy Division, University of California, Los Angeles, CA 90095-1547, USA (malkan@astro.ucla.edu).}

\shortauthor{Spinoglio, Tommasin \& Malkan}
\shorttitle{AGN and starburst: from IR photometry to IR spectroscopy}

\listofauthors{L. Spinoglio, S. Tommasin \& M.A. Malkan}
\indexauthor{Spinoglio, L.}
\indexauthor{Tommasin, S.}
\indexauthor{Malkan, M.A.}

\abstract{Infrared photometry and later infrared spectroscopy provided 
powerful diagnostics to distinguish between the main emission mechanisms 
in galaxies: AGN and Starburst. After the pioneering work on infrared 
photometry with \textit{IRAS} in the far-IR and the S.Pedro Martir and 
ESO ground-based work in the near-IR, ISO photometry extended up to 
200$\mu$m the coverage of the galaxies energy distributions. 
Then \textit{Spitzer} collected accurate mid-infrared spectroscopy 
on different samples of galaxies. We will review the work done on 
the 12$\mu$m galaxy sample since the times of \textit{IRAS} photometry 
to the new \textit{Spitzer} spectroscopy. The main results on the 
multifrequency data of 12$\mu$m selected Seyfert galaxies are 
presented and discussed in the light of unification and evolution models. 
The spectroscopic work of \textit{Spitzer}  will soon be complemented 
at longer wavelengths by the \textit{Herschel} spectrometers and in 
the future by \textit{SPICA} at higher redshift.}

\addkeyword{Galaxies: Active}
\addkeyword{Galaxies: Starbursts}
\addkeyword{Infrared: Galaxies}

\begin{document}
\maketitle

\section{Introduction}
\label{sec:intro}
The interrelationship between \textit{star formation} 
and  \textit{accretion} onto massive black holes is crucial to understanding
galaxy formation and evolution. On a cosmic scale, the evolution
of supermassive black holes 
appears tied to the evolution of the star-formation rate 
\citep{mar,mer}. 
The growth of bulges through \textit{star formation} 
may be directly linked to the growth
of black holes through accretion \citep{hec04}. On a local scale, evidence is
mounting that \textit{star formation} and nuclear activity are linked. Two possible
evolutionary progressions can be predicted: HII/Starburst galaxies $\rightarrow$ Seyfert\,2
\citep{sto, kau}, or a fuller 
scenario of HII/Starburst  galaxies $\rightarrow$ Seyfert\,2 $\rightarrow$
Seyfert\,1 \citep{hun99, lev01, kro}.
These predict that galaxy interactions, leading to the
concentration of a large gas mass in the circumnuclear region of
a galaxy, trigger starburst emission. Then mergers and
bar-induced inflows can bring fuel to a central black hole,
stimulating AGN activity. While relatively young ($\sim$ 1 Gyr)
stellar populations are found in more than half of Seyfert 2s \citep{sch, gond, rai},
they are also found in broadlined AGNs \citep{kau}. 
However, any firm conclusion cannot rely on optical
spectra of optically selected samples of galaxies. Photometric
mid-IR studies  \citep{emr, maio} 
did indeed find that Seyfert 2s galaxies 
more often have enhanced star formation than Seyfert 1s and the near and far-IR observations
of the 12$\mu$m galaxy sample \citep{s95} show systematic differences between type 1's 
and type 2's spectral energy distributions. However,
detailed \textit{infrared spectroscopy} is  necessary to better separate the star formation and accretion components in the energy budget of active galaxies and strengthen the hypothesis 
of an evolutionary difference between different types of active galaxies.
Better understanding the  \textit{star formation} versus  
\textit{accretion} connection requires mid-infrared
spectroscopy of representative samples of active galaxies in the
local universe. 
This is because active galactic nuclei (AGNs) are
often very dusty, locally and even at high redshifts 
\citep{pri,be03}. 
Similarly, \textit{star formation} activity,
which often coexists with AGN activity, is also heavily
enshrouded in dust. 

We review in this article the great amount of observational work that has been done on the 12$\mu$m galaxy sample and in particular on its subsample of active galaxies. 

\section{The 12$\mu$m active galaxies sample}
\label{sec:sample}

\begin{figure}[!b]
  \includegraphics[width=\columnwidth]{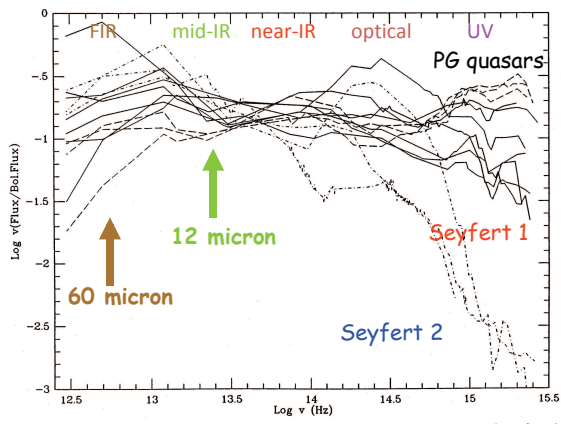}
  \caption{Spectral energy distributions 
    of 13 AGN normalized to the bolometric fluxes 
  \citep{sm89}.}
  \label{fig:12um}
\end{figure}

\begin{figure}[!t]
  \includegraphics[width=\columnwidth]{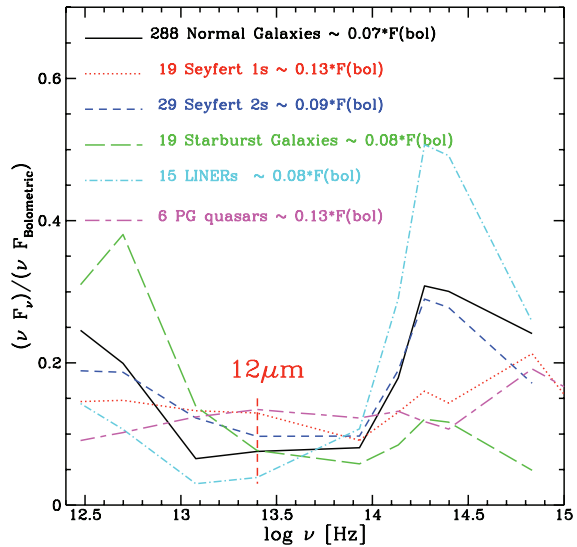}
  \caption{Average SEDs
  of the various types of galaxies normalized 
  to their bolometric fluxes  \citep{s95}.}
  \label{fig:seds}
\end{figure}

The sample that is less biased and most representative
of the local active galaxies populations
is selected from the 12$\mu$m Galaxy Sample (12MGS), an IRAS-selected
all-sky survey flux-limited to 0.22 Jy at 12$\mu$m 
(Rush, Malkan \& Spinoglio 1993, hereafter RMS)
and form the complete sub-samples of Seyfert 1s and Seyfert
2s of the entire 12MGS. This is essentially a bolometric
flux-limited survey outside the galactic plane, because of the
empirical fact that all {\it galaxies} emit a constant
fraction of their total bolometric luminosity at 12$\mu$m. This
fraction is $\sim$ 9-13\% for AGNs 
and $\sim$ 7-8\% for normal and starburst galaxies, independent of star
formation activity \citep{s95}. For Seyfert galaxies and quasars, 
this can be seen in Figure~\ref{fig:12um}, which shows the spectral energy distributions of 
13 active galaxies normalized to their bolometric flux: the minimum 
scatter among the different types of galaxies appears in the range 
7-12$\mu$m \citep{sm89}. For the different types of 12$\mu$m 
selected galaxies, normal, Seyfert, starburst galaxies and LINERs, 
compared to a small sample of PG quasars, the spectral energy 
distributions normalized to their bolometric fluxes are shown in Figure~\ref{fig:seds}.  
Finally, the 12MGS is
less subject to contamination by high star-formation rate objects
than other infrared samples defined at longer
wavelengths \citep{hun99}.

\begin{figure}[!h]
  \includegraphics[width=\columnwidth]{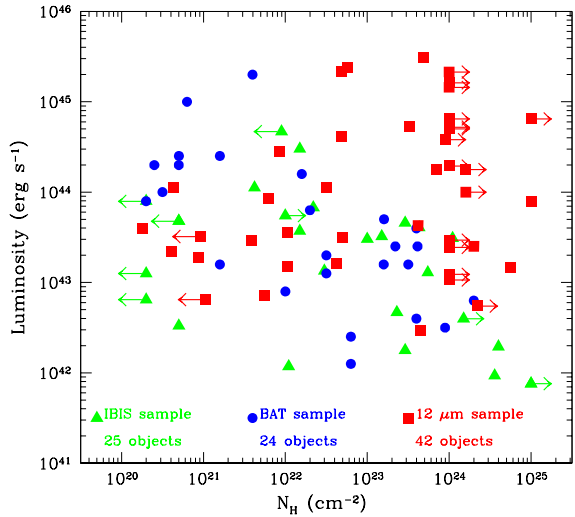}
  \caption{Luminosity-Hydrogen absorption column density \textit{plane} of the 12$\mu$m Seyfert (squares) detected at hard X-rays. Hard-X-ray selected AGNs, from IBIS (triangles) and Bat (circles), are shown for comparison.}  
 \label{fig:NH}
\end{figure}

12 $\mu m$ selection finds obscured objects via re-radiation of their primary emission by dust. 
An alternative way of finding obscured AGN is by selecting directly on their accretion 
radiation at hard X-rays, 
insensitive to all but the heaviest intrinsic absorption. 
However, unlike the 12MGS, as shows Figure~\ref{fig:NH}, 
the hard X-ray selected samples miss most of the Compton thick objects ($N_{\rm H} > 10^{24}$~cm$^{-2}$).

The 12MGS contains 53 Seyfert 1s and 63 Seyfert 2s (RMS). 
This sample has a complete set of observations at virtually every 
wavelength: full IRAS and near-IR coverage (RMS;
Spinoglio et al 1995),  X rays \citep{r96}, 
optical spectroscopy, radio \citep{rme}, %
optical/IR imaging \citep{hun99, h99}, 
100-200$\mu$m far-infrared photometry from ISOPHOT 
\citep{s02}. 
In the recent years 10$\mu$m imaging \citep{gor}, 
2.8-4.1$\mu$m slit spectroscopy \citep{ima3,ima4}
optical spectropolarimetry \citep{tra1,tra3}
radio observations \citep{the0,the1} 
and \textit{Spitzer} low resolution spectra \citep{buc} 
have been collected for most of the Seyfert galaxies in our sample.
Finally a first article on the high resolution \textit{Spitzer} spectra has been 
published \citep{t08} and another one is completing the mid-infrared spectral 
coverage of almost all the sample \citep{t09}.

\subsection{The 12$\mu$m and the line luminosity functions of Seyfert galaxies}

\begin{figure}[!h]
  \includegraphics[width=\columnwidth]{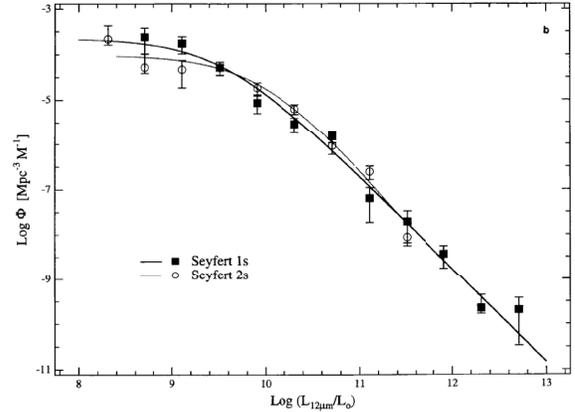}
  \caption{ The 12$\mu$m luminosity function of Seyfert galaxies (RMS).}
  \label{fig:12um_lf}
\end{figure}

The 12$\mu$m luminosity function of the Seyfert galaxies of the 12MGS has 
been derived by RMS. There is no significant difference between the two types of Seyfert, as one
can see from Figure~\ref{fig:12um_lf}, except for the fact that at the highest luminosities ($L>10^{12}  L_{\odot}$)
only type 1's are found. We note,  however, that this can be due 
to the inclusion in this latter class of the few blazars and quasars present
in the 12MGS.

\begin{figure}[!h]
  \includegraphics[width=\columnwidth]{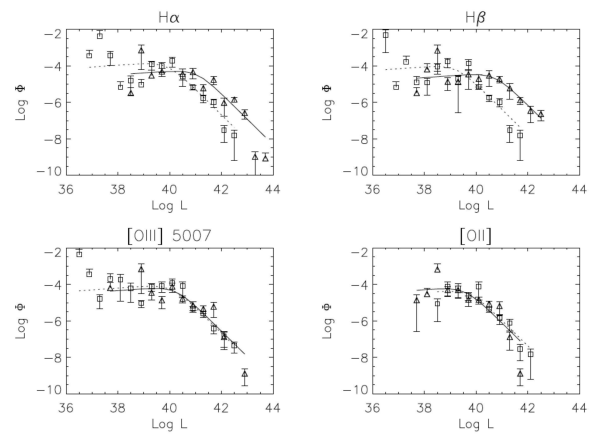}
  \caption{ Emission line luminosity functions for H$\alpha$, H$\beta$, [OIII]$\lambda$5007\AA~ and   [OII]$\lambda$3727\AA~ for the 12$\mu$m selected Seyfert galaxies \citep{r09}. Triangles represent Seyfert 1's and squares Seyfert 2's.}
  \label{fig:line_lf}
\end{figure}

The analysis of the optical and ultraviolet emission line
spectra has been done recently \citep{r09}.
The luminosity functions of all the narrow lines for which there is enough statistics 
([OI]$\lambda$6300\AA, [OII]$\lambda$3727\AA, [OIII]$\lambda$4959\AA, [OIII]$\lambda$5007\AA, [NII]$\lambda$6584\AA~ and [SII]$\lambda$$\lambda$6717+6734\AA)
are the same for both types of Seyfert's. 
Only the H$\alpha$ and H$\beta$ luminosity functions 
show more Seyfert 1's at high luminosities compared to Seyfert 2's.
This difference can be understood because type 1's have substantial emission
in both these lines from their Broad Line Regions.
The luminosity functions for H$\alpha$, H$\beta$,
[OIII]$\lambda$5007\AA~ and [OII]$\lambda$3727\AA ~are displayed in Figure~\ref{fig:line_lf}.

The agreement between the 12$\mu$m continuum and the narrow lines luminosity functions testifies that 
these quantities are all isotropic and are not affected by the geometry or disk/torus orientation. This also implies that the 12$\mu$m selection is not biased against or in favor of Seyfert types 1 or types 2. 

\section{Infrared photometry}
\label{sec:ir-phot}
\subsection{Total near-infrared fluxes}

\begin{figure}[!t]
  \includegraphics[width=\columnwidth]{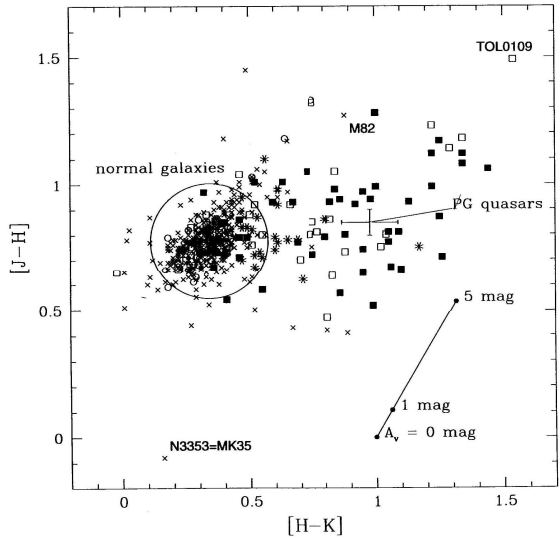}
  \caption{[J - H] \textit{vs.} [H - L] 
  diagram with Seyfert 1's (\textit{filled squares}), Seyfert 2's (\textit{open squares}), 
  Normal's (\textit{crosses}) \citep{s95}.}
  \label{fig:jhk}
\end{figure}

\begin{figure}[!b]
  \includegraphics[width=\columnwidth]{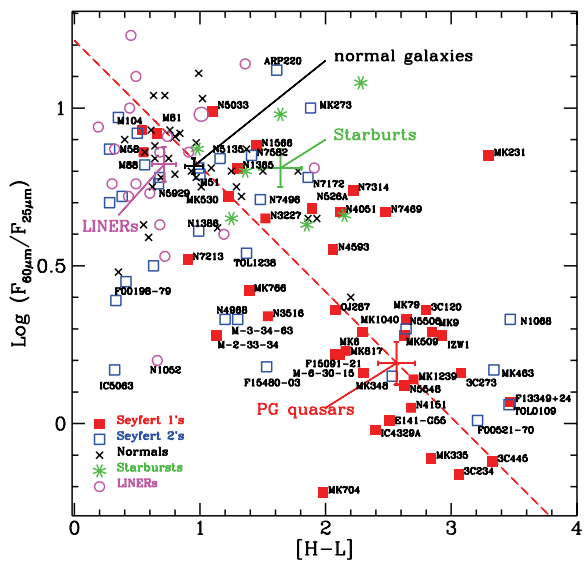}
  \caption{Composite near-far-IR color-color diagram: the [H - L] color versus the 60$\mu$m/25$\mu$m
flux ratio \citep{s95}.}
  \label{fig:col_col1}
\end{figure}

Infrared photometry has been used to isolate star formation and accretion processes.
A large observational effort has been done to observe in the J, H, K and, in some 
cases, L bands as many as 321 galaxies from
the 12$\mu$m galaxy sample \citep{s95}. At the 2.1m telescope of the S. Pedro 
Martir Observatory, thanks to a collaboration 
with Luis Carrasco and Elsa Recillas, we observed the northern galaxies, while those of the 
southern hemisphere were observed at the 1m ESO telescope (La Silla, Chile). 

To be able to compare the large beam IRAS data (RMS)
with the near-IR data, derive spectral energy distributions and build 
meaningful color-color diagrams, we have computed the near-IR total fluxes using growth curves \citep{s95}.
Combining the IRAS and near-IR data we were able to derive the spectral energy distributions
(SED) of hundreds of galaxies. As presented in the previous section, we  show in Figure~\ref{fig:seds}
the average SEDs, normalized to their bolometric flux, of normal, Seyfert and starburst galaxies, as well as LINERs and PG quasars.

We have found that color-color diagrams are very effective in separating the different galaxy types. 
The usual [J - H] versus [H - K] color-color diagram shown in Figure~\ref{fig:jhk} can separate many Seyfert galaxies from normal galaxies: at 2.2$\mu$m the strong thermal emission from hot dust grains illuminated by the active nucleus rises above starlight from red giants, peaking in the H photometric band and dominating the stellar component in galaxies.   

\begin{figure}[!b]
  \includegraphics[width=\columnwidth]{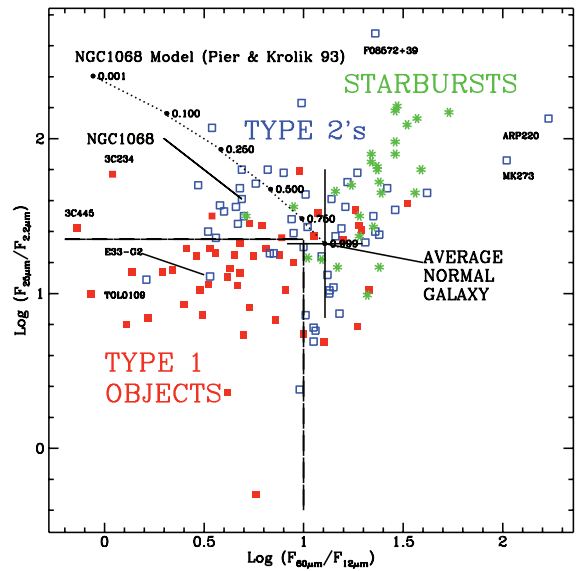}
  \caption{Color-color diagram separating Seyfert 1's, 2's and starburst galaxies.
  The dotted line shows a mixture of the NGC1068 
  torus model  colors 
   \citep{pk3} and the average galactic colors. The symbols are the same as in the previous figure \citep{s95}.}
  \label{fig:col_col2}
\end{figure}

In Figure~\ref{fig:col_col1} we show the [H - L] color versus the 60$\mu$m/25$\mu$m flux ratio. This plot can separate normal and starburst galaxies 
from Seyfert type 1 and PG quasars, because the former have steep far-IR slopes and bluer near-IR spectra, while the latter have flatter far-IR SEDs and redder near-IR slopes. As most Seyfert 1's are located at the lower right part of the diagram, while most type 2's lie in the upper left corner, where only a few Seyfert 1's are found (most of which are nearby Messier galaxies) demonstrate that the SEDs of the two Seyfert populations are indeed different.

In Figure~\ref{fig:col_col2} we show the 60$\mu$m/12$\mu$m versus the 25$\mu$m/2.2$\mu$m, as suggested previously for an analysis of the CfA Seyfert galaxies \citep{emr}, where it is clear a segregation of the Seyfert type 1's in the lower left part of the diagram, at the flatter spectral slopes: 30/48 type 1s and only 5/52 type 2s occupy this region. The presence of an optically thick edge-on torus  \citep{pk3} would indeed steepen the SED by absorbing high energy radiation and re-emitting it at longer wavelengths.

\subsection{Bolometric luminosities of Seyfert and normal galaxies}

Spinoglio et al (1995) determined for the first time the bolometric luminosities 
of a large sample of galaxies, both Seyfert and normal galaxies. One of their
most important result is the linear relation found between the 12$\mu$m and
\textit{bolometric} luminosities for both Seyfert and normal galaxies. 

\begin{figure}[!h]
  \includegraphics[width=\columnwidth]{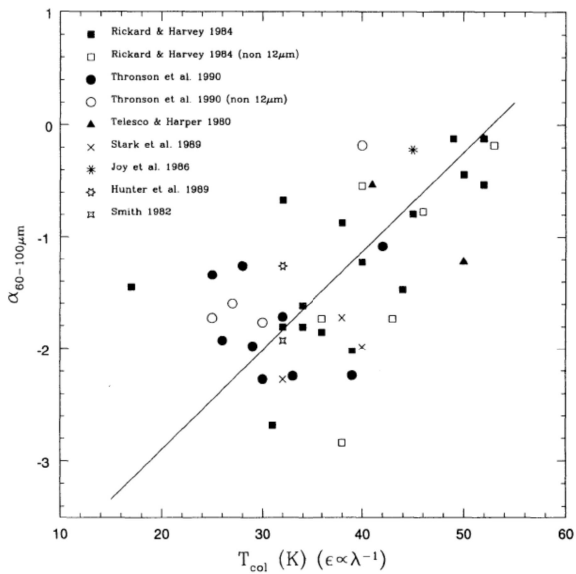}
  \caption{Least squares fit of the spectral index $\alpha_{60-100\mu m}$ as a function of the color
  temperature, assuming graybody emission 
  \citep{s95}.}
  \label{fig:bol_cor}
\end{figure}

To obtain reliable \textit{bolometric luminosities} for the galaxies, we have used the IRAS [60 -100$\mu$m] 
color to predict the far-IR turnover. Figure~\ref{fig:bol_cor} shows the correlation between color temperature and IRAS spectral index. From the fit shown, we derived the relation: 
$$T_{color} = 11.4 \times (\alpha_{60-100\mu m} + 4.67)K$$ 
We then computed the submillimeter fluxes beyond 100$\mu$m assuming a graybody 
at the derived color temperature, with
a dust emissivity $\epsilon \propto \lambda^{-1}$.

\begin{figure}[!h]
  \includegraphics[width=\columnwidth]{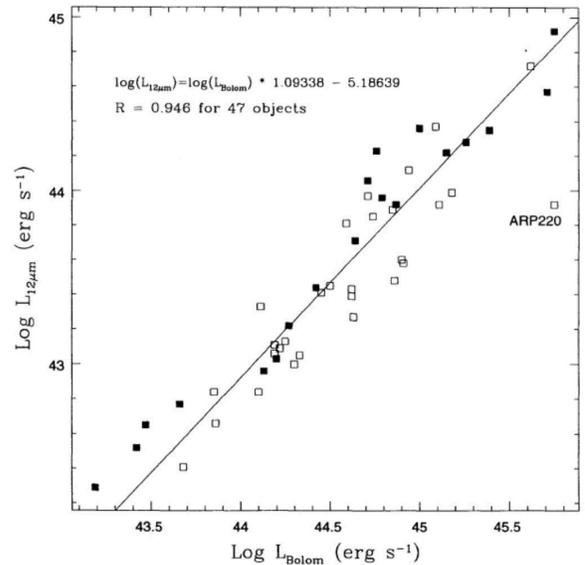}
  \caption{12$\mu$m 
  \textit{vs} bolometric luminosity for Seyfert 1's (\textit{filled squares}) and 2's  (\textit{open squares}).
  The line represents the least squares fit to all data, except Arp220 
  \citep{s95}.}
  \label{fig:12um_bol_sy}
\end{figure}

\begin{figure}[!t]
  \includegraphics[width=\columnwidth]{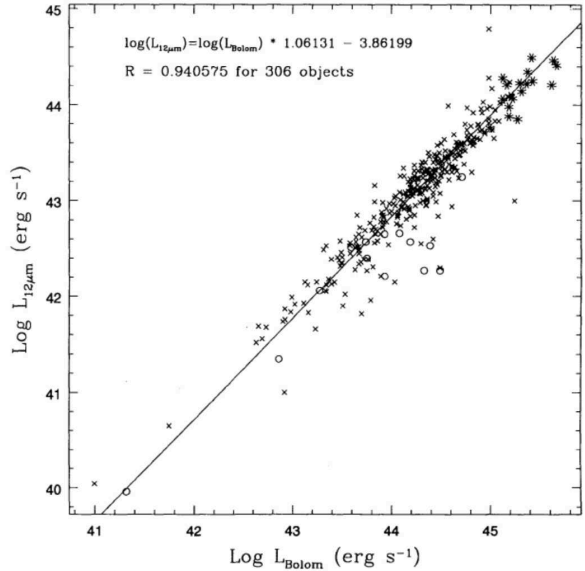}
  \caption{12$\mu$m 
  \textit{vs} bolometric luminosity for normal galaxies (\textit{crosses}), starburst's  (\textit{asterisks}) and LINERs (\textit{open circles}).
  The line represents the least squares fit to all data 
  \citep{s95}.}
  \label{fig:12um_bol_no}
\end{figure}

For the Seyfert galaxies, the relation between the 12$\mu$m and
the bolometric luminosity is shown in Figure~\ref{fig:12um_bol_sy}, with a slope of 1.09
(with a regression coefficient of R=0.95). 
We notice that a similar relation between bolometric luminosity and the 
other IRAS bands monochromatic luminosities is poorer compared to the 
12$\mu$m band: at 25$\mu$m the slope 
is 1.19 (R=0.90), at at 60$\mu$m the slope is 1.19 (R=0.91) and at 
100$\mu$m the slope is 1.13 (with a poorer R=0.93). 

For the normal galaxies, the relation between the 12$\mu$m and
the bolometric luminosity is shown in Figure~\ref{fig:12um_bol_no}, with a slope of 1.06
(with a regression coefficient of R=0.94). 
A similar relation between bolometric luminosity and the 
other IRAS bands monochromatic luminosities is poorer compared to the 
12$\mu$m band: at 25$\mu$m the slope 
is 1.15 (R=0.93), at at 60$\mu$m the slope is 1.12 (R=0.94) and at 
100$\mu$m the slope is 1.11 (with a poorer R=0.89).

\subsection{Extending to 200$\mu$m with ISO}

\begin{figure}[!b]
  \includegraphics[width=\columnwidth]{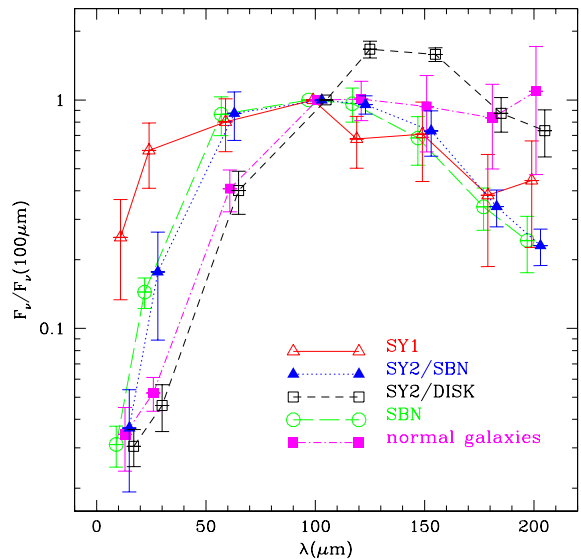}
  \caption{The average spectral energy distributions of Seyfert, 
  starburts and normal galaxies, normalized to the 100$\mu$m flux \citep{s02}.}
  \label{fig:iso_seds1}
\end{figure}

With the launch of the \textit{Infrared Space Observatory}, we 
were able to collect far-infrared photometry between 100 and 200$\mu$m 
for the 12$\mu$m galaxies \citep{s02}. 
We have followed Rowan-Robinson \& Crawford (1989)
 in using the 12-25-60-100 $\mu$m colors
to identify those galaxies in our sample that closely
resemble the SEDs of the \textit{quiescent cirrus} disk, the starburst
component, and the \textit{pure Seyfert} nucleus. For each
of the three types of galaxies - the normal spirals, the starburst
galaxies, and the Seyfert 1's - we have selected those
objects lying in the two IRAS color-color diagrams close
(i.e., within 0.2 mag) to the colors of \textit{pure disk}, \textit{starburst},
and \textit{Seyfert} components \citep{rrc}.
We have plotted
in Figure~\ref{fig:iso_seds1} the average 12-200$\mu$m SEDs for each class. A strong distinction
is apparent out to 200$\mu$m between the quiescent disk
component and the starburst component. These components represent 
the extremes of minimal and
maximal recent star formation found in the least and most
luminous galaxies, respectively.
The pure Seyfert spectrum is rather similar to the pure starburst spectrum
between 100 and 200$\mu$m. Both show a relative lack
of cold dust, and the SeyfertsÕ spectra tend to be weaker at
120$\mu$m.

\begin{figure}[!b]
  \includegraphics[width=\columnwidth]{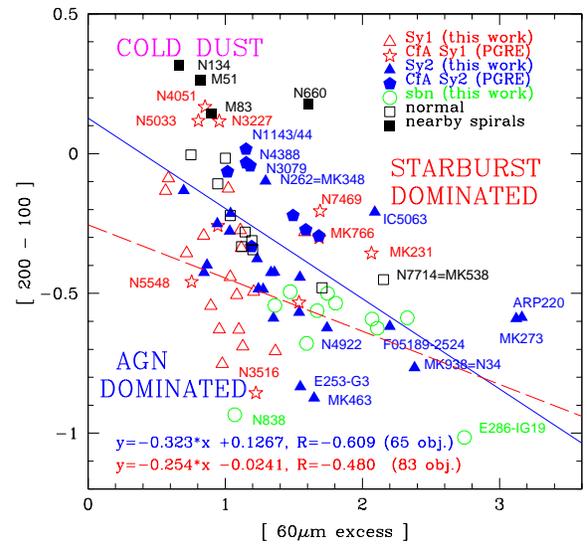}
  \caption{[200 - 100$\mu$m] color vs. the 60$\mu$m excess diagram of galaxies belonging to
  the 12$\mu$m galaxy sample \citep{s02}.}
  \label{fig:excess}
\end{figure}
The Seyfert 2's are spread all around the IRAS color-color
diagrams. As can be seen in Figure~\ref{fig:iso_seds1}, some of them (SY2/SBN) have
IRAS spectra close to the pure starburst template. 
And indeed, their ISOPHOT far-infrared
spectra also match the pure starburst spectrum
well, since their infrared continuum appears to be dominated
from dust around star-forming regions. Those Seyfert
2's with IRAS colors like quiescent cirrus disks (SY2/DISK) also resemble
pure disks in the 100-200$\mu$m region. Again, it appears that
the Seyfert 2 nucleus contributes a minor fraction of the
observed far-infrared luminosity in those objects.

We have chosen as the indicator of enhanced recent star
formation, which warms dust around HII regions, the 60$\mu$m \textit{excess} as the ratio of the observed 60$\mu$m  flux to the flux that a source would have at 60$\mu$m from power-law
interpolation of the flux between 12 and 100$\mu$m. Figure~\ref{fig:excess} shows
the [200 - 100$\mu$m] color versus the 60$\mu$m \textit{excess}.

While this diagram does
not separate the galaxies of different classes perfectly, it
nevertheless shows that they cluster preferentially in different
regions of the diagram. Seyfert 1's (excluding six objects
of the CfA sample) cluster in a no 60$\mu$m excess region with
a color [200 - 100$\mu$m] $<$ 0. The starburst galaxies cluster in the
central area of the diagram and have all 60$\mu$m excess. Normal
galaxies and nearby spirals have no 60$\mu$m excess
(except two objects) and a [200 - 100$\mu$m] color between -0.5 and +0.5.
Seyfert 2's are widely spread all over the diagram, but with a
60$\mu$m excess generally higher than Seyfert 1's.

We suggest that the diagram shown in Figure~\ref{fig:excess} can be
used to separate the starburst-dominated objects from the
AGN-dominated ones. Objects located in the upper right
part of the diagram are likely \textit{starburst-dominated}, while
those at the left, having a fainter excess, are the \textit{AGN-dominated}
objects. We suggest that starburst activity in galaxies,
i.e., with high rates of current star formation, results in
excess emission in the 60$\mu$m band accompanied by a general
heating of the galactic interstellar medium and thus a
decrease of the [200 - 100$\mu$m] color.


\section{Infrared spectroscopy}
\label{sec:ir-spec}


Mid-IR and far-IR spectroscopy of fine-structure emission lines
are powerful tools to understand the physical conditions in
galaxies. 

\begin{figure}[!h]
  \includegraphics[width=\columnwidth]{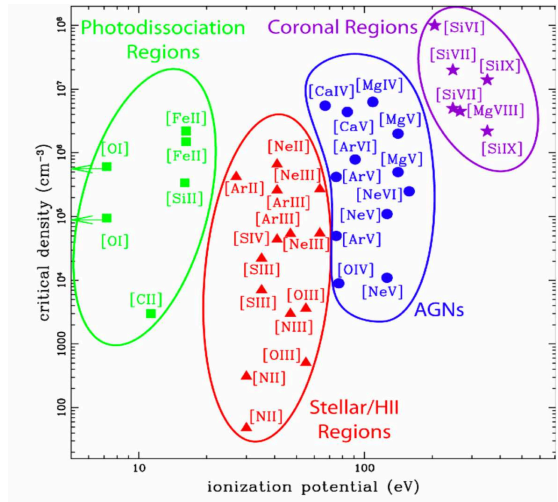}
  \caption{The power of the infrared fine-structure lines to trace the different physical regimes: 
  \textit{From left to right:} low-ionization/photodissociation regions, stellar/HII regions,
  AGN emission line regions, Coronal line regions \citep{sm92}.}
  \label{fig:irspec}
\end{figure}

Figure~\ref{fig:irspec} shows the critical density (i.e. the density for which the
rates of collisional and radiative de-excitation are equal) of
a line versus the ionization potential of its ionic
species. It shows how these lines can measure
density and ionization of the gas: the ratio
of lines with similar critical density, but different
ionization potential, measures the ionization,
while the ratio of lines with similar ionization potential, but
different critical density, measures the density \citep{sm92}.
Lines from different emission regions in
galaxies are shown with different symbols. 
Infrared spectroscopy has a thorough
diagnostic power for gas with densities from 10$^2$ cm$^{-3}$ 
to 10$^8$ cm$^{-3}$ and ionization potentials up to 350 eV, 
using the so called coronal lines. 
Moreover, increasing its wavelength, an IR spectral line becomes
more insensitive to dust extinction, and can therefore probe regions 
highly obscured at optical or even near-to-mid infrared wavelengths.

Besides the ionic fine-structure lines, the mid-IR spectrum of 
galaxies also contains strong features due to the emission of Polycyclic 
Aromatic Hydrocarbons \citep{pl}, 
hereafter PAH. These features have been observed in ultraluminous IR galaxies with
the \textit{ISO} SWS spectrometer \citep{ge98}. They are present while star 
formation is active and disappear when illuminated by the 
strong ionizing field of active nuclei.

\subsection{Far-IR spectroscopy with \textit{ISO-LWS} }

\begin{figure}[!t]
  \includegraphics[width=\columnwidth]{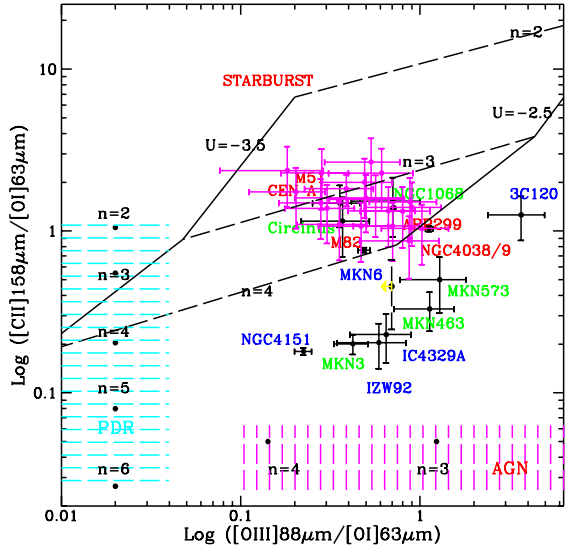}
   \caption{ISO-LWS line ratio diagram  \citep{spi03}. 
   Seyfert 1's: blue, Seyfert 2's: green, starburst galaxies: red. Pink crosses: 
   nearby galaxies \citep{ne01}. Grid: starburst models with different densities 
   and ionization parameter. Vertical hatched area: the [CII]/[OI] ratio for PDR 
   models (log n=2-6 cm$^{-3}$, log G$_{0}$=3). Horizontal hatched area: the [OIII]/[OI] 
   ratio for AGN models (log U=-2.5, log n= 3, 4 cm$^{-3}$).  }
  \label{fig:lws}
\end{figure}

Far-infrared spectroscopy has been so far collected by the LWS spectrometer onboard of 
\textit{ISO} \citep{kes96} only on a few bright active and ultraluminous IR galaxies, 
showing an unexpected sequence of features from strong [OIII]52, 88 $\mu$m and
[NIII]57 $\mu$m line emission to detection of only faint
[CII]157$\mu$m line emission and [OI]63 $\mu$m in absorption, and molecular lines almost always in absorption \citep{fis}.
A few studies have been dedicated to 
individual galaxies, e.g. M82 \citep{co99}, Arp220 \citep{g04}, NGC1068 \citep{s05}, Mrk231 \citep{g08}. 
A systematic far-infrared spectroscopic survey of Seyfert and ULIRGs will have to wait for the \textit{Herschel}
mission. However the few data available already revealed the diagnostic power of the FIR fine structure lines.
Figure~\ref{fig:lws} \citep{spi03} 
shows  the [CII]158$\mu$m/[OI]63$\mu$m ratio versus the [OIII]88$\mu$m/[OI]63$\mu$m ratio. 
Normal galaxies cluster together with the Seyfert's having strong starburst emission (e.g. NGC1068) 
and coincide with predicted ratios for typical starburst galaxies. However, at lower values of the 
[CII]/[OI] ratio, most of the AGN are clustering in a strip with higher ratio of [OI]/[CII] 
compared to starburst galaxies. This may arise from 
X-Ray Dissociation Regions (XDRs), whose [OI]/[CII] ratios are larger than in PDRs.
Thus the 3 strongest FIR emission lines can separate the 3 basic energy
sources in galaxies: 
a) the AGN produces strongly emission from highly ionized gas, with [OIII] being prominent
in the NLR, but also unusually strong [OI]63$\mu$m because this neutral line is strong in
XDRs  \citep{mei07},
and has a very high critical density ($\sim 10^6 cm^{-3}$). 
All the classical Seyfert galaxies in Figure~\ref{fig:lws} lie in the bottom part of the diagram,
below any of the starburst models.
b) recent star formation, which produces [OIII] in the high-excitation HII regions,
as well as strong [CII]158$\mu$m in the PDRs 
which tend to surround these star forming regions; and
c) pure PDR emission from the quiescent disk of the spiral galaxy, which produces strong
[CII] and [OI] emission, but no [OIII].  The most quiescent spirals lie in the upper left
side of the diagram \citep{ne01}.

\subsection{Mid-IR spectroscopy with \textit{Spitzer} }

In the very rich mid-infrared spectra of active galaxies we can identify various indicators of
\textit{AGN dominance}, e.g. the line ratios of [NeV]/[NeII], [NeV]/[SiII], [OIV]/[NeII], as well as 
indicators of \textit{star formation dominance}, e.g. the PAH emission bands, the H$_2$ rotational lines
and nebular emission lines mainly originated in HII regions, e.g. from [SIII] and [NeII].

\begin{figure}[!h]
  \includegraphics[width=\columnwidth]{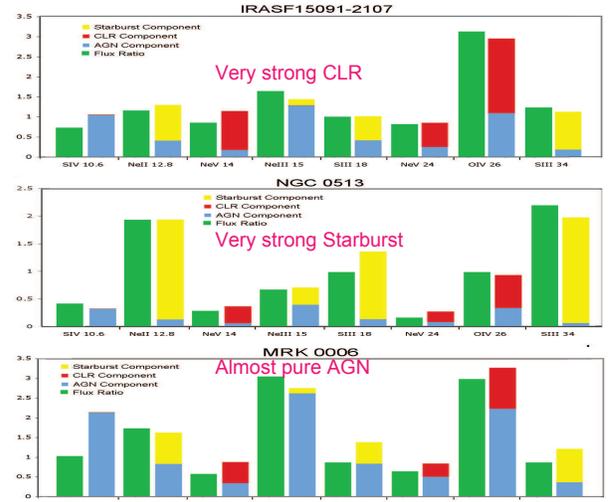}
  \caption{ Comparison of the observed and modelled spectra for three sample Seyfert galaxies:
 \citep{k09}.}
  \label{fig:kevin}
\end{figure}

The diagnostic power of the mid-IR fine structure lines can be quantified from the decomposition of 
the observed spectra in terms of three different photoionized components \citep{k09}: 
(a) \textit{AGN model}: with metallicity
Z solar, spectral slope $\alpha$=-1.7, density n=10$^3$, and ionization potential: 
$logU=-2$ \citep{gro04};
(b) \textit{Starburst model}: with metallicity $Z=2 \times Z_{\odot}$, 
$age~range=0.1-6 Myr$, $log R=(M_{cl}/M_{\odot})/(Po/k) = -6$, \citep{dop06}; 
 (c) \textit{a Coronal line region model} (CLR): $log U$ = 0 at the inner radius of the region
 and spectral slope $\alpha$= -1.0 \citep{sm92}.
Figure~\ref{fig:kevin} shows such a decomposition for the mid-IR spectra of three sample objects: 
IRASF15091-2107 for which
a strong CLR is present, NGC513 for which a strong Starburst component is necessary to fit the data and
Mrk6, which is almost a "pure" AGN.

\begin{figure}[!b]
  \includegraphics[width=\columnwidth]{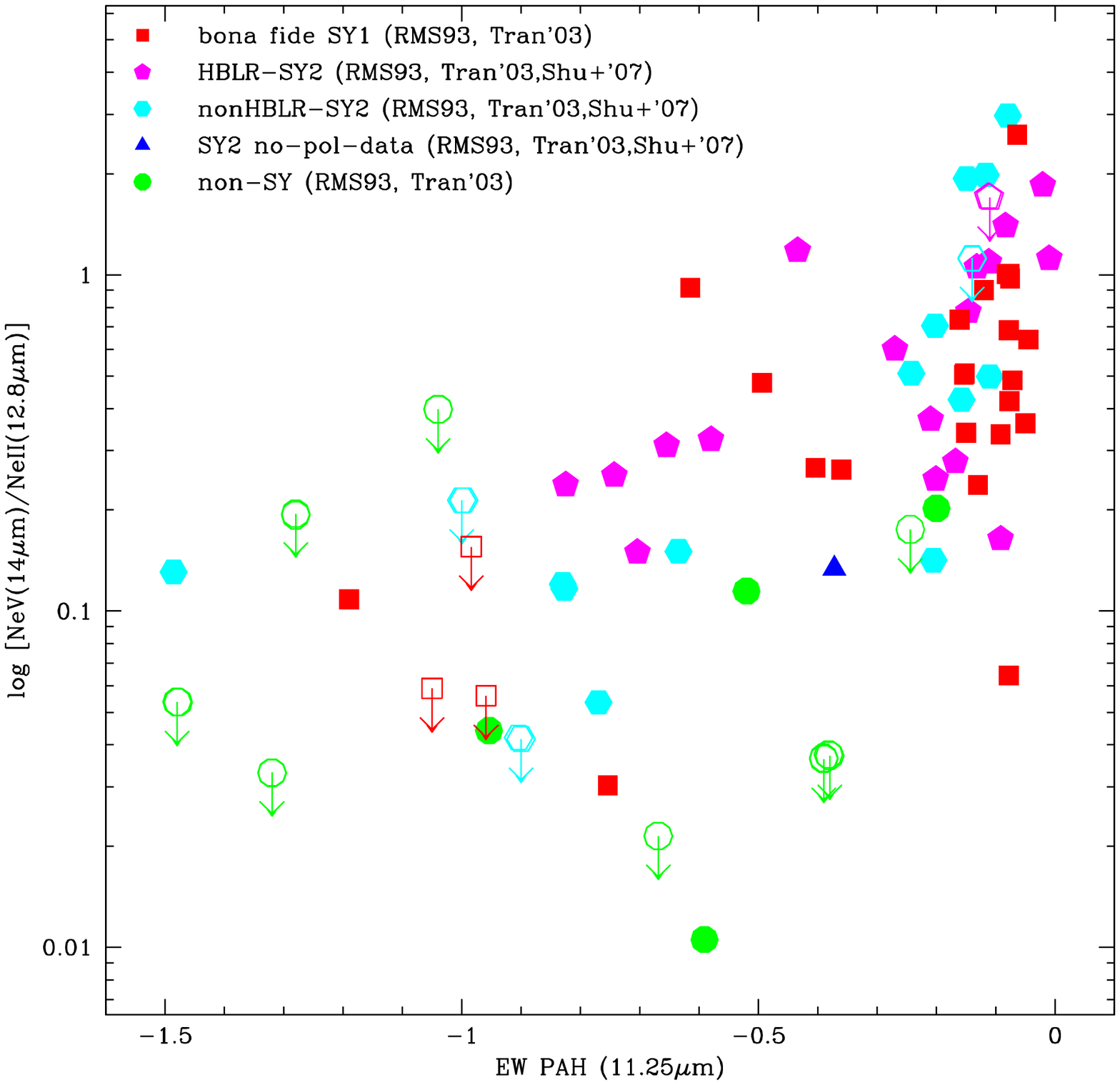}
  \caption{  [NeV]14.3$\mu$m/[NeII]12.8$\mu$m line ratio 
versus the equivalent width of the 11.25$\mu$m PAH.
 \citep{t08,t09}.}
  \label{fig:neon_pah}
\end{figure}

The first results of the \textit{Spitzer} spectroscopic survey of the Seyfert galaxies of the 12$\mu$m 
sample \citep{t08} show a clear inverse trend between the indicator of 
\textit{AGN dominance}, the [NeV]14.3$\mu$m/[NeII]12.8$\mu$m line ratio, 
and the equivalent width of the 11.25$\mu$m PAH feature, which can be 
considered as an indicator of the \textit{star formation dominance}, as shown in Figure~\ref{fig:neon_pah}.
Here the Seyfert galaxies have been reclassified, following the results of spectropolarimetry \citet{tra1,tra3}, 
in type 1's (including the classical Seyfert 1's and the hidden Broad Line Region Seyfert 2's, as
discovered through spectropolarimetry) and ''pure" type 2's (for which a BLR was not detected). 
Most of the type 1 objects, 
including both Seyfert 1s and hidden Broad Line Region Seyfert 2s, are
located at high values of the [NeV]14.3$\mu$m/[NeII]12.8$\mu$m line ratio
and very low or absent PAH emission.

\begin{figure}[!t]
  \includegraphics[width=\columnwidth]{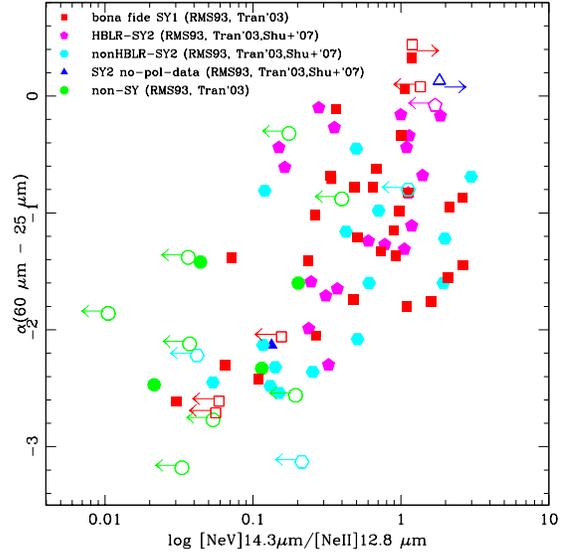}
  \caption{ The mid-to-far-IR spectral index $\alpha_{(60\mu m-25\mu m)}$ versus the 
[NeV]14.3$\mu$m/[NeII]12.8$\mu$m line ratio.
\citep{t08}.}
  \label{fig:alpha_neon}
\end{figure}

Another diagnostic diagram using both spectroscopic and photometric results is shown in Figure~\ref{fig:alpha_neon}: 
the spectral index between 25 and 60$\mu$m $\alpha_{(60\mu m-25\mu m)}$ versus the 
[NeV]14.3$\mu$m/[NeII]12.8$\mu$m line ratio. A clear trend shows that when the  
\textit{AGN dominance} increases, the spectral index flattens. Most of type 1 objects
appear to be concentrated in the upper right part of the diagram, at high values of 
\textit{AGN dominance} and flat mid-to-far-IR slopes.

\begin{figure}[!h]
  \includegraphics[width=\columnwidth]{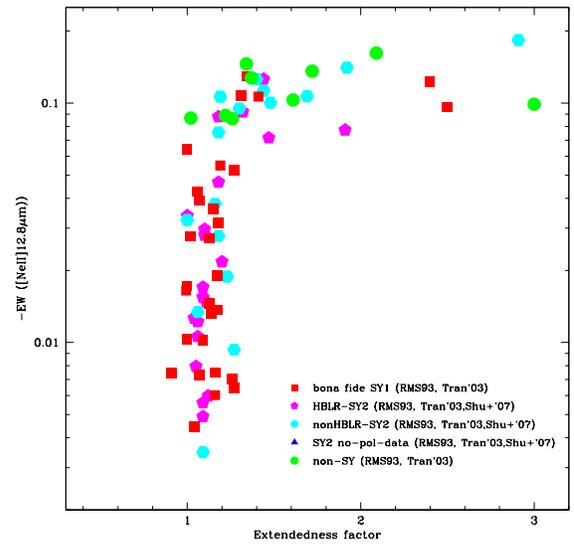}
  \caption{  [NeII]12.8$\mu$m line equivalent width as a function of the source extendednes \citep{t08}.}
  \label{fig:ext_ne2}
\end{figure}

The two channels of the Spitzer high-resolution spectrometer: 
SH 9.6-19.5$\mu$m with slit size 4.7$\arcsec$ $\times$ 11.3$\arcsec$ and LH 19-39$\mu$m 
with slit size 11.1$\arcsec$ $\times$ 22.3 $\arcsec$
allow Òmulti-aperture photometryÓ in the overlapping parts (19.0-19.5$\mu$m).
The ratio of the flux measured in LH to that measured in SH gives the ÒextendednessÓ of the source.
We used this measure of the extendedness of the source to estimate the line emitting regions \citep{t08}.
In Figure~\ref{fig:ext_ne2} we plot the [NeII]12.8$\mu$m line equivalent width as a function of the source extendedness.
We notice that those sources showing a significant mid-IR extendedness 
are type 2 objects or non-Seyfert galaxies and have the highest 
[NeII]12.8$\mu$m line equivalent width. An high [NeII]12.8$\mu$m line equivalent width is
a measure of a strong star formation component. This is not the case for the high excitation lines, 
originated from the AGN, such as [NeV] and [OIV], for which no apparent trent appears between 
source extendedness and line EWs  \citep{t08}.


\section{Analysis of the 12$\mu$m sample multi-frequency dataset}
\label{sec:ana}

\begin{figure}[!t]
  \includegraphics[width=\columnwidth]{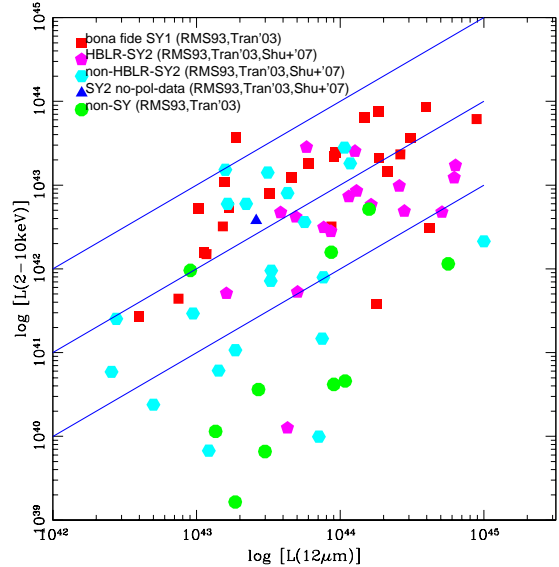}
  \caption{ Corrected (unabsorbed) X-ray (2-10 keV) luminosity as a function of the 12$\mu$m luminosity. The three lines
  from the top to the bottom indicate the loci of L(2-10keV) = L(12$\mu$m) (\textit{upper});  
  L(2-10keV) = 0.1 $\times$ L(12$\mu$m) (\textit{middle}); and L(2-10keV) = 0.01 $\times$ L(12$\mu$m) (\textit{lower}), which are used in the text to roughly separate the different objects.}
  \label{fig:lx_l12}
\end{figure}

The 12MGS has been observed extensively from the radio to the X-rays and we can use the large set
of data to search for correlations between different observed quantities. 
To show an example, we want to relate the X-ray luminosity, measuring the accretion, to the bolometric
luminosity, as given by the  12$\mu$m luminosity.
We plot in Figure~\ref{fig:lx_l12} 
the \textit{unabsorbed} 2-10keV luminosity and the 12$\mu$m luminosity.

Following the finding of Spinoglio \& Malkan (1989) and Spinoglio et al (1995) 
that the 12$\mu$m luminosity is
linearly proportional to the \textit{bolometric} luminosity, at a given L$_{bol}$ 
in Figure~\ref{fig:lx_l12} a sequence can be identified with decreasing accretion luminosity: from 
Seyfert 1's $\rightarrow$ HBLR-Seyfert 2's  $\rightarrow$ \textit{pure} Seyfert 2's.
Although these results are to be considered preliminary, as no statistical method has
yet been applied, 
most Seyfert 1's have: 

$0.1 \times L(12{\mu}m) < L(2-10keV) < L(12{\mu}m)$

\noindent Most HBLR-Seyfert 2's have: 

$0.01 \times L(12{\mu}m) < L(2-10keV) < 0.1 \times L(12{\mu}m)$

\noindent Most pure Seyfert 2's and non-Seyfert's have: 

$L(2-10keV) < 0.01 \times L(12{\mu}m)$

We preminilarily suggest that black hole accretion, as measured by X-rays, is the dominant mechanism
determining the observational nature of a galaxy: when accretion is not an important energy source, we
have galaxies without Seyfert nuclei, dominated by stellar evolution processes (called here non-Seyfert's), then when accretion increases we have a sequence from
the \textit{pure} Seyfert 2's, to the HBLR-Seyfert 2's and finally when  accretion dominates the bolometric luminosity,
we have the Seyfert 1's.

\begin{figure}[!b]
  \includegraphics[width=\columnwidth]{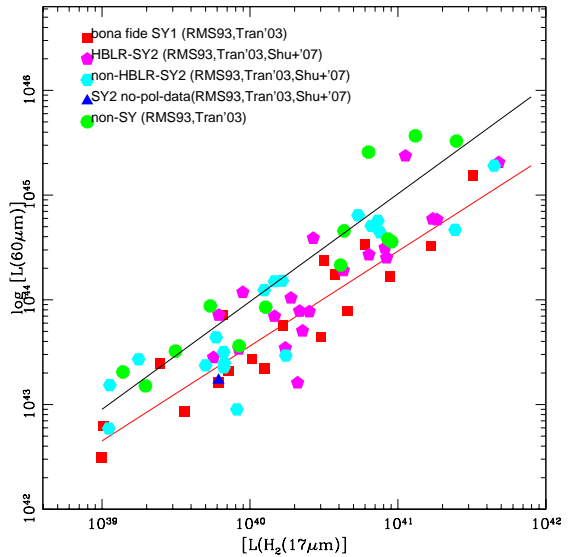}
  \caption{ Total 60$\mu$m luminosity as a function of the H$_2$ 17$\mu$m line luminosity.
  The two lines from the top to the bottom are least squares fits of the non-Seyfert galaxies and of the 
  Seyfert 1's, respectively (see the text)}
  \label{fig:lh2_l60}
\end{figure}

In an analogous way, we try to correlate the IRAS 60$\mu$m luminosity (measuring the integrated star formation activity) 
and the H$_2$ S(1) line luminosity (typical star formation indicator) in Figure~\ref{fig:lh2_l60}.

\noindent Most Seyfert 1's and HBLR-Seyfert 2's have: 

$L(H_2) \sim 5 \times 10^{-4} \times L(60{\mu}m)$

\noindent Most pure Seyfert 2's and non-Seyfert have: 

$L(H_2) \sim 10^{-4} \times L(60{\mu}m)$


If we make least squares fits to the two extreme populations of Seyfert type 1's and non-Seyfert
galaxies, we obtain a sequence of two almost parallel 
lines of the form $Log(L(60{\mu}m)) = a \times log(L(H_2)) + b $ from the bottom to the top: 
\begin{itemize}
\item[-] for Seyfert 1's: $a$=0.905, $b$=7.325, with a regression coefficient of R=0.928;
\item[-] for non-Seyfert's: $a$=1.030, $b$=2.797, with R=0.925.
\end{itemize}

There are two interpretations of this behavior: either the more active galaxies (type 1's) have enhanced
H$_2$ emission \citep{rig02}, or at a given H$_2$ luminosity, type 2's (and non-Sy) have $L(60{\mu}m)$ 5 
times higher than type 1's, because of an enhanced star formation process.



\section{Spectroscopy of higher redshift galaxies with Herschel \& SPICA}

To understand how the two processes of black hole accretion and 
star formation shared the energy budget during
galaxy evolution, we need to separate these two processes along the 
history of galaxies, and -to do this- rest-frame near-to-mid infrared 
spectroscopy is needed on galaxies as a function of their redshift.
We predicted the line intensities of Seyfert and starburst galaxies at 
increasing redshift, considering the ISO spectra of three local template 
objects: NGC1068 \citep{ale,s05}, 
the prototypical Seyfert 2 galaxy, containing both an AGN and a starburst; 
NGC6240 \citep{lu03}, a bright starburst with obscured AGN and  
M82 \citep{fo01,co99}, the prototypical starburst galaxy. 
 We then computed the line intensities as a function of redshift 
 (in the range z=0.1-5), assuming that the line luminosities scale 
 as the bolometric luminosity and that there is a luminosity evolution 
 proportional to the (z+1)$^2$, consistent with the Spitzer results at 
 least up to redshift z=2 \citep{per05}. 

\begin{figure}[!h]
  \includegraphics[width=\columnwidth]{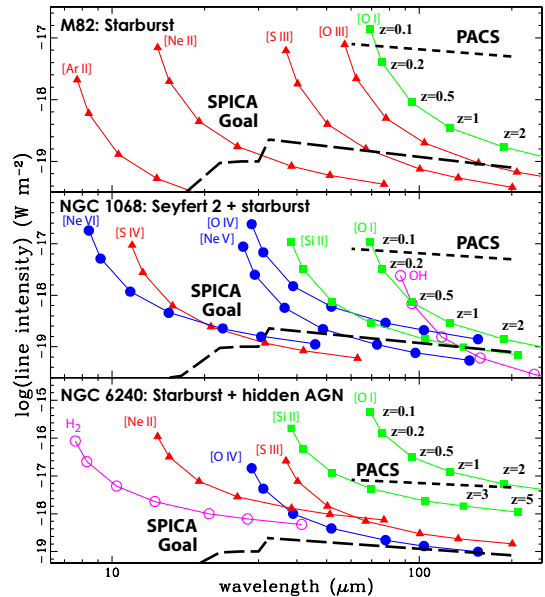}
  \caption{ Line observability with PACS onboard of \textit{Herschel} and  with \textit{SPICA}.  Few selected 
  diagnostic lines are shown as a function of redshift for the three template objects M82, NGC1068 and NGC6240   
  (from the top to the bottom). The lines have the same symbols as in Figure~\ref{fig:irspec}, except for the
  open circles, representing molecular lines. 
  Line intensities are given in W m$^{-2}$. The short dashed and long-dashed lines give the  5$\sigma$, 
  1 hour sensitivities of the PACS and SPICA spectrometers.}
  \label{fig:templates}
\end{figure}

For simplicity\footnote{ We note that the
dependence on different cosmological models is not very strong.
The popular model with $\Omega_{M}$= 0.27, $\Omega_{vac}$=0.73,
H$_{0}$=71 km s$^{-1}$ Mpc$^{-1}$ shows greater dilutions,
increasing with z, by factors of 1.5 for z=0.5 to 2.5 for z=5. In
this case the line intensities of Figure~\ref{fig:templates} would decrease by these
factors.}, we adopted an Einstein-De Sitter model Universe, with
$\Omega_{\Lambda}$ = $\Omega_{vac}=0$ and $\Omega_{M}$= 1,
H$_{0}$=75 km s$^{-1}$ Mpc$^{-1}$. The luminosity distances have
been derived using:
$$d_{L} (z)= (2c/H_{0})\cdot [1+z - (1+z)^{1/2}]$$
The results for the three template objects are reported in a graphical form in 
Figure~\ref{fig:templates}, where the intensities of selected lines are plotted 
as a function of the redshift. Among the brightest lines are shown the 
[SIV]10.5$\mu$m, the [NeII]12.8$\mu$m and the [OIII]52$\mu$m diagnostic
for the stellar/HII regions;  the [NeV]24.3$\mu$m and the [OIV]25.9$\mu$m, 
for the AGN component, the [OI]63$\mu$m and the [SiII]33.5$\mu$m, 
for the photodissociation regions and the OH and H$_2$ rotational lines for the
warm molecular component.

The 5$\sigma$, 1 hour sensitivities of the PACS spectrometer onboard of \textit{Herschel} 
and of the two spectrometers foreseen at the focal plane of the JAXA mission \textit{SPICA} 
(Space Infrared Telescope for Cosmology and
Astrophysics) \citep{na04,sw08} are shown in the figure for comparison.

It is clear from the figure that the PACS spectrometer will be able to observe only the most 
favorable object (NGC6240) up to z=2 in the brightest line ([OI]63$\mu$m), while the SPICA 
spectrometers \textit{goal} sensitivities will allow deep infrared spectroscopic studies 
for all templates at z $\sim$ 1-2 for most lines and at z even higher for the brightest lines.


\smallskip
\smallskip
\smallskip
\smallskip

\noindent \textit{Acknowledgements: This work has been funded in Italy by the Italian Space Agency (ASI).}


\end{document}